\documentclass{iaus}
\usepackage{graphicx}

\def\ms{M$_{\odot}$}

\def\dexkpc{~dex~kpc$^{-1}$}

\title[~~Properties and chemical evolution of M31]
{Properties and Chemical Evolution of the M31 Disk: Comparison with
the Milky Way }
\author[Hou et al.]
{J.L. Hou$^1$, J. Yin$^{1,4}$, S. Boissier$^{2}$, N. Prantzos$^3$,
R.X. Chang$^1$, L. Chen$^1$ \\ }

\affiliation{$^1$Shanghai Astronomical Observatory, CAS, 80 Nandan
Road, Shanghai, 200030, China \\email: {\tt houjl@shao.ac.cn,crx@shao.ac.cn,
chenli@shao.ac.cn} \\[\affilskip]
$^2$Laboratoire d'Astrophysique de Marseille, BP8, Traverse du
Siphon, 13376 Marseille, Cedex 12, France  \\email: {\tt samuel.boissier@gmail.com } \\[\affilskip]
$^3$Institute d'Astrophysique de Paris, CNRS, 98bis Arago 75014,
Paris, France \\email: {\tt prantzos@iap.fr} \\[\affilskip]
$^4$Graduate School, the Chinese Academy of Sciences, Beijing, 100039, China
\\email:{\tt jyin@shao.ac.cn}}

\pubyear{2008}
\volume{254}  
\jname{The Galaxy Disk in Cosmological Context} \editors{J. Andersen
(Chief Editor), J. Bland-Hawthorn \& B. Nordstrom}

\begin{document}

\maketitle

\begin{abstract}

By adopting the chemical evolution model of the Milky Way disk, we
have studied the star formation and chemical evolution history for
M31 galaxy disk. We mainly concentrated on the global properties of
the M31 disk. The model has been scaled to the related disk
parameters, mainly the disk scale length and total disk baryon mass
of M31, which we have adopted to be $r_d$=5.5kpc and $M_{tot} =
7\times 10^{10}$ \ms. It is found that, when the classical Kennicutt
star formation law was applied, the obtained radial profiles of gas
surface density and star formation rate (SFR) have great difference
from the observed results in M31 disk. Then we have adopted modified
SFR as we did for the Milky Way galaxy, that is the SFR is radial
dependent. Detailed calculations show that by adjusting the star
formation efficiency, it is possible to get reasonable gas and
abundance profiles, but the total disk SFR is a factor of 2-3 higher
than that estimated from observations. And also the predicted SFR
radial profile is also much higher than what GALEX observed in the
outer part. Possible reasons could be that the M31 disk has been
interacted by other factors which seriously altered the star
formation history, or the observed SFR is underestimated due to
inappropriate extinction correction.



\end{abstract}

\firstsection 

\section{Introduction}

As one of the three disk galaxies in the Local Group, Andromeda
galaxy (M31, or NGC224) provides an unique opportunity for testing
theory of galaxy formation and evolution (Widrow et al. 2003; Renda
et al. 2005; Widrow \& Dubinski 2005; Brown et al. 2006; Tamm et al.
2007; Tempel et al. 2007). In order to understand how star formation
has influenced the evolution of M31 disk, Williams (2003a, 2003b)
has measured star formation history in several regions of the M31
disk from the KPNO/CTIO Local Group Survey. It is found that the
total mean star formation rate for the disk is about 1 \ms
yr$^{-1}$. By deep HST photometry, Bellazzini et al. (2003) have
studied the metallicity distributions and star formation history in
many locations of the disk (see also Ferguson \& Johnson 2001;
Sarajedini \& van Duyne 2001; Olsen et al. 2006).

However, some basic properties of the Milky Way and M31 galaxies are
still uncertain despite years of detailed observations and analysis.
For example, the total masses of two galaxies could be varied by a
factor of few based on the current published literatures in which
the mass of galaxy is mostly derive from kinematic analysis of the
galaxy and its neighbors. And there also exists disagreement over
which of the two galaxies should be more massive (Evans \& Wilkinson
2000; Gottesman et al. 2002).

In this work, we apply an infall model, which is similar to what has
been used for the Milky Way galaxy, to M31 galaxy. We will mainly
concentrate on the global properties of the M31 disk based on the
currently available data. We would like to know whether the similar
model could explain both the Milky Way and M31 disk observations. If
not, what is the main difference in the formation history of those
two disks. We would like to demonstrate and hope to understand the
similarities and differences of star formation and chemical
evolution history between M31 and the Milky Way disks.

\section{Observed properties of the M31 and Milky Way disks}

{\bf Disk Scale Length and Mass: } We assume the disk has an
exponential total surface density now (unit: $M_{\odot}pc^{-2}$):

\begin{equation}\label{eq:sigma}
    \Sigma_{tot}(r,t_g)=\Sigma_0(0,t_g)e^{-r/r_d}
\end{equation}
$r_d$ is the disk scale length. $\Sigma_0(0,t_g)$ is the central
surface density at the present time. The disk scale length $r_d$ has
some complexity since it is wavelength dependent. It is obtained on
the basis of surface brightness profiles in various bands. In table
1, we list all the available observed disk scale lengths and scaled
to the same distance of 785 kpc (McConnachie et al. 2005). It can be
found that the overall values are consistent for different bands,
except for the shorter wavelengths which are likely to be affected
by dust extinctions.

\begin{table}[!t]
\noindent Table 1. M31 exponential disk scale length \\
[2mm]
\begin{tabular}{lcccl}
\hline \hline
 band & R$_d$         &   D   &   R$_d$     & Refs. \\ [1mm]
      & (kpc)         & (kpc) & (785kpc)      &       \\
\hline
 U   & 6.8$\pm$0.4    &  690     &  7.7      &  (1)  \\
 B   & 5.8$\pm$0.3    &  690     &  6.6      &  (1)  \\
 V   & 5.3$\pm$0.3    &  690     &  6.0      &  (1)  \\
 R   & 5.2$\pm$0.3    &  690     &  5.9      &  (1)  \\
     & 5.4$\pm$0.13   &  784     &  5.4      &  (2)  \\
 I   & 5.6$\pm$0.4    &  770     &  5.7      &  (3)  \\
 K   & 4.2$\pm$0.4    &  690     &  4.8      &  (4)  \\
 L   & 6.08$\pm$0.09  &  783     &  6.1      &  (5)  \\ [1mm]
\hline \hline
\end{tabular} \\ [1mm]
Refs: (1) Waterbos \& Kennicutt 1988; (2) Geehan et al. 2006; (3)
Worthey et al. 2005; (4) Hiromoto et al. 1983; (5) Barmby et al.
2006.
\end{table}

In this work we shall adopt an averaged value from four observed
values from three bands (R, I, K), which is $r_d$=5.5 kpc.

For the Milky Way disk, its exponential disk scale length has been
widely measured (see a review of Hammer et al. 2007). The generally
accepted value is about r$_d$ = 2.3 kpc (R or I band, Hammer et al.
2007 ). This is less than half of the M31 disk.

For the total mass of M31 disk, observational data and some mass
models have also given some rough estimations. In their
disk-bulge-halo model, Widrow et al. (2003) shown that the best
model requires the M31 disk mass be about $7\times 10^{10}$\ms.
Recent mass model of Geehan et al.(2006) also give a similar disk
mass value about $7.2\times 10^{10}$\ms by adopting the disk
mass-to-light ratio of 3.3. In this paper, we will adopt the M31
total disk mass to be $M_{tot} = 7\times10^{10}$\ms.

In short summary, we can know that M31 disk is about 2 times massive
than the Milky Way disk ($3.5\times 10^{10}$\ms) and the scale
length is 2.4 times longer than that of the Milky Way disk (2.3kpc).

{\bf Gas and SFR Profiles :} Thanks to the GALEX, we are now also
able to get star formation rate radial profile for a number of local
galaxies derived from the UV (not H$\alpha$ data) continuum
(Boissier et al. 2007). In general, the correlation between star
formation rate and gas surface density is compatible with empirical
Kennicutt (1998) type law with some scatter in the low surface
density end. But as Boissier et al. (2007) show this correlation for
some individual galaxies, especially for M31, is quite abnormal (see
their Fig.6).

\begin{figure}[!t]
  \centering
  \includegraphics[height=6cm,width=10cm]{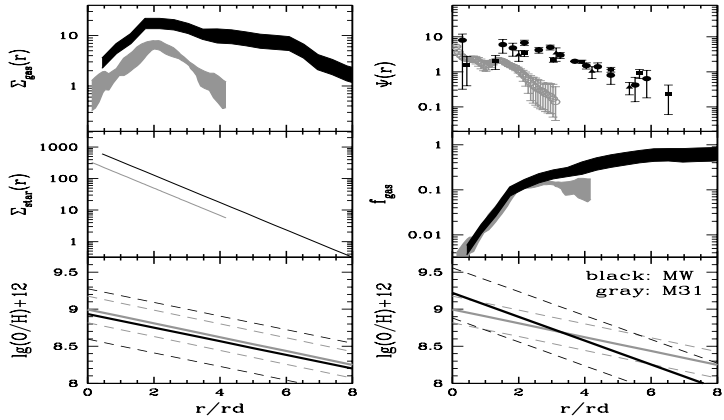}\\
    \caption{Observed profiles for Milky Way and M31 disks, where
    the radius are scaled by scale length for each disk. For the
    abundance gradient of the Milky Way disk, two sets of slope values
    are plotted, one is $-$0.07 \dexkpc from Rudolph et al. (2006)
    (lower right panel), the other is $-$0.040 \dexkpc from Deharveng et al.
    (2000) and Dalfon \& Cunha (2004)(left lower panel). The dashed lines
    are the typical scatter of the observations. }
  \label{Fig:scaleobs}
\end{figure}

The observed profiles of gas and SFR for both Milky Way and M31 are
plotted in the upper two panels of Fig~\ref{Fig:scaleobs}, where we
have scaled the radius to disk scale length $r_d$.

{\bf Disk Abundance Gradients : } In the lower two panels of
Fig~\ref{Fig:scaleobs}, we plot the radial abundance gradients for
the two disks. The M31 gradient data are taken from different
literatures (Blair 1982; Dennefeld \& Kunth 1981; Trundle et al.
2002). It is evident that the M31 disk abundance ($\sim -0.017$
\dexkpc, ) is about 4 times smaller than that of the Milky Way disk
($-0.07$ \dexkpc, Rudolph et al. 2006). If we express them in scale
length (dex/$r_d$), then the scaled gradient of M31 disk is about 2
times smaller (-0.09 vs. -0.16, right lower panel in
Figure~\ref{Fig:scaleobs} ).

However, there are some observations indicate that the abundance
gradient along the Milky Way diks could be shallower than the
commonly adopted results. For example, based on detailed
observations of 34 Galactic HII regions between 5 and 15 kpc,
Deharveng et al. (2000) reported a value of oxygen abundance
gradient about $-0.039\pm0.005$ \dexkpc. A shallow gradient is also
obtained by Daflon \& Cunha (2004), who derived a mean abundance
gradient about $-0.042\pm0.007$ \dexkpc for a sample of 69 members
of 25 open clusters, OB associations and HII regions with
Galactocentric distances between 4.7 and 13.2 kpc. Recently, Chen \&
Hou (2007) have updated their open clusters sample by adding more
objects, and also find that the disk iron abundance gradient for all
clusters is about $-0.058\pm0.006$ \dexkpc, smaller than their
previous results. For young clusters, the gradient could be more
smaller. In this case, the gradients per scale length for two disks
are similar (left of lower panels of Fig~\ref{Fig:scaleobs}).

\section {A Unified Description of the Milky Way and M31 disk}

Observations have provided a number of data for both Milky Way and
M31 galaxies. There are clear indications of similarities and
differences between M31 and Milky Way galaxy, from bulges, disks to
halos (Wyse 2002). From figure.~\ref{Fig:scaleobs}, several points
can be drawn:

(1) The Milky Way has more extended gas and star distributions
relative to scale length. The Milky Way disk has more gas and higher
star formation rate also. Especially in the outer part of the disk.

(2) The distribution of scaled gas fraction between two disks are
quite similar. This is consistent with the situation that Milky Way
has more gas and higher SFR. However, the overall gas fractions of
the two disks are quite different, in which the Milky Way has twice
as much as that of the M31 disk.

(3) The scaled abundance gradients between two disks are similar if
the we adopt the smaller value of gradient for Milky Way disk.

It can be seen, when the observed profiles are scaled to scale
length, two disks show some similar properties. The main differences
come from the SFR profiles in the outer part of two disks.
Therefore, it is instructive to see if we could establish a unified
chemical evolution model to reproduce the observed properties for
both disks.

\section{The Model Results}

In this section, we give brief discussions about the model results.
Details about the model can be found in Yin et al. (2008), and also
in Boissier \& Prantzos (1999, 2000) and Hou et al. (2000).

The main parameter in our model is the SFR. In Boissier \& Prantzos
(1999,2000), the SFR was expressed as a modified Kennicutt-Schmidt
law (hereafter M-KS law):
\begin{eqnarray} \label{eq:sfrBP00}
\Psi(r,t) = \alpha \Sigma_{gas}^{n} r^{-1}
\end{eqnarray}
where $\Sigma_{gas}$ is in unit of $M_{\odot}pc^{-2}$, $r$ is in
unit of kpc. The index $n$ was chosen to be $n=1.5$ on an empirical
basis. And For the Milky Way disk, $\alpha$ =1.0. They adopted this
M-KS law in subsequent models for external spirals and the models
can successfully reproduce global properties of spirals ( Boissier
\& Prantzos 2000, 2001; Boissier et al. 2001).

In the following calculations, we will adopt M-KS law and with
$\alpha$ being a free parameters for Milky Way and M31 disk.

\begin{figure*}[!t]
  \centering
  \includegraphics[angle=-90,width=10cm]{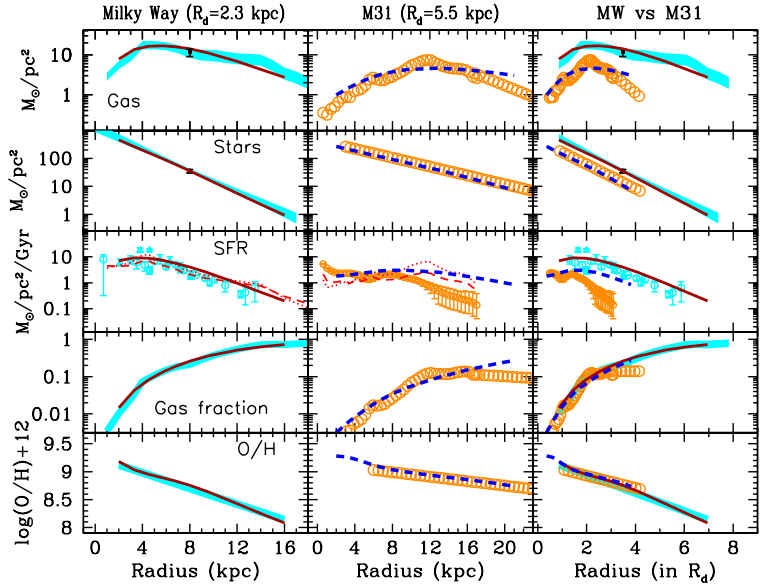}\\
  \caption{Current profiles of Gas, star, gas fraction, SFR and oxygen abundance
  predicted by the model. $\tau(r) = 2.5r/r_d$. $\alpha_{MW} = 1.0,
  \alpha_{M31} =2.0$}
  \label{Fig:Profiles}
\end{figure*}

In Fig.~\ref{Fig:Profiles}, we show the model predictions of the
radial profiles for gas, star, SFR and oxygen abundance and compare
with the observations. The stellar profile is obtained by
subtracting the observed gas profile from the exponential disk
(governed by the disk scale length and the central mass surface
density). The star formation efficiency (parameter $\alpha$) is
adjusted in order to get a satisfactory present day gas profile in
both Milky Way and M31 disks. For the Milky Way disk, it is well
justified that the acceptable value is about $\alpha \sim 1.0 $ ,
while for M31, we need a value of $\alpha \sim 2.0$, which is twice
as much as that for the Milky Way disk, in order to get satisfied
results. In the third column, we plot the model and observed results
by scaling the radius to scale length.

Observationally, M31 has flatter gradient than the Milky Way disk.
But there are also observations show that the Milky Way has flatter
gradient as we mentioned before. If the Milky Way has a gradient
about $-0.04$ \dexkpc, then the scaled values for M31 and Milky Way
are quite consistent.

Model problems for the M31 disk come from the inconsistency between
the predicted and observed SFR profiles. No matter how we adjust the
model parameters, it is unable to get reasonable M31 SFR profile,
especially in the outer disk, where the model will predict much
higher SFR in order to get lower gas content.

This difficulty could be partially solved by changing the star
formation history of the disk. If star formation efficiency were
higher in the past, then more gas would be consumed and resulting
less gas content at present. This scenario is at least partially
supported by observations. Recent observations based on deep surveys
have provided enormous information concerning star formation
history, morphology, metallicity and kinematics in various regions
of the M31 disk, bulge and halo (Ibata et al. 2005; Ferguson et al.
2005; Guhathakurta et al. 2005; Brown et al. 2006; Olsen et al.
2006). It is generally accepted that the M31 galaxy has experienced
many more small mergers or a few large ones during its evolution
history. Such kind of mergers should not destroy the M31 disk, but
must have great influence on disk properties, including disk SFR
history and gas evolution.

\section{Summary}

In this work, we have studied the chemical evolution for the M31
disk based on the model similar to the Milky Way disk (Boissier \&
Prantzos 1999, 2000). The main observational constrains for M31 disk
are the profiles of gas and star formation rate, the abundance
gradients. The star formation profile of the M31 disk is from recent
UV data of GALEX (Boissier et al. 2007).

We tried to use a unified chemical evolution model to predict the
disk properties of both M31 and Milky Way by using the modified KS
SFR law. When the results are scaled to disk scale lengths, we find
that the M31 need a higher star formation efficiency than the Milky
Way disk. And the model always predicts a higher SFR in the outer
part of the M31 disk. It is also found that the adopted different
infall time scales are not very sensitive to the radial profiles of
gas, SFR and abundances. The inside-out disk formation is mainly
controlled by the radial dependence of star formation rate instead
of the time scale.

\acknowledgements This work is supported by national science
foundation of China with No.10573028, No.10573022 and by 973 program
with No. 2007CB815402.

\def\cjaa{ChJAA}
\def\apj{ApJ}
\def\apjl{ApJL}
\def\apjs{ApJS}
\def\aj{AJ}
\def\aap{A\&A}
\def\araa{ARA\&A}
\def\aapss{A\&AS}
\def\mnras{MNRAS}
\def\nature{Nature}
\def\apss{Ap\&SS}
\def\pasp{PASP}

\section*{References}
\smallskip
\parindent=0pt
\everypar{\hangindent 1.2cm} \footnotesize


Barmby P., Ashby M.L.N., Bianchi L. et al., 2006, ApJ, 650, L45

Bellazzini M., Cacciari C., Federici L. et al., 2003, A\&A, 405, 867

Blair W.P., Kirshner R.P., Chevalier R.A., 1982, ApJ, 254, 50

Boissier S., Boselli A., Prantzos N. et al., 2001, MNRAS, 321, 733

Boissier S. \& Prantzos N., 1999, MNRAS, 307, 857

Boissier S. \& Prantzos N., 2000, MNRAS, 312, 398

Boissier S. \& Prantzos N., 2001, MNRAS, 325, 321

Boissier S., Gil de Paz A., Boselli A. et al., 2007, ApJS,173, 524

Brown T.M., Smith E., Ferguson H.C. et al., 2006, ApJ, 652, 323

Chamcham K., \& Tayler R.J., 1994, MNRAS, 266, 282

Chen L. \& Hou J.L., 2007, IAU Symposium 248

Daflon S. \& Cunha K., 2004, ApJ, 617, 1115

Deharveng L., Pe\~na M., Caplan J. et al., 2000, MNRAS, 311, 329

Dennefeld M., \& Kunth D., 1981, AJ, 86, 989

Evans N. W., \& Wilkinson M. I., 2000, MNRAS, 316, 929

Ferguson A. M. N., \& Johnson R. A., 2001, ApJ, 559, 13

Ferguson A. M. N., Johnson R. A., Faria D. C. et al., 2005, ApJ,
622, L109

Geehan J. J., Fardal M. A., Babul A. et al., 2006, MNRAS, 366, 996

Gottesman S.T., Hunter J.H., Boonyasait V., 2002, MNRAS, 337, 34

Guhathakurta P., Gilbert K. M., Kalirai J. S. et al., 2005, BAAS,
37, 1386

Hammer F., Puech M., Chemin L., Flores H. \& Lehnert M. D., 2007,
ApJ, 662, 322

Hiromoto N., Maihara T., Oda N. et al., 1983, PASJ, 35, 413

Hou J. L., Prantzos N., Boissier S., 2000, A\&A, 362, 921

Ibata R., Chapman S., Ferguson A. M. N. et al., 2005, ApJ, 634, 287

Kennicutt R. C., 1998, ARA\&A, 36, 189

McConnachie A., Irwin M., et al., 2005, MNRAS, 356, 979

Olsen K. A. G., Blum R. D., Stephens A. W. et al., 2006, AJ, 132,271

Renda A., Kawata D., Fenner Y. et al., 2005, MNRAS, 356, 1071

Rudolph A. L., Fich M., Bell G. R. et al., 2006, ApJS, 162, 346

Sarajedini A., \& van Duyne J. 2001, AJ, 122, 2444

Tamm A., Tempel E. \& Tenjes P., 2007, arXiv:0707.4375v1

Tempel E., Tamm A. \& Tenjes P., 2007, arXiv:0707.4374v1

Trundle C., Dufton P. L., Lennon D. J. et al., 2002, A\&A, 395, 519

Walterbos R. A. M., \& Kennicutt R. C., 1988, A\&A, 198, 61

Widrow L.M., Perrett K.M., Suyu S. H., 2003, \apj, 588, 311

Widrow L.M., \& Dubinski J., 2005, \apj, 631, 838

Williams B.F., 2003a, AJ, 126, 1312

Williams B.F., 2003b, MNRAS, 340, 143

Worthey G., Espa\~na A., MacArthur L.A. et al., 2005, ApJ, 631, 820

Wyse R.F.G., 2002, European Astrophysical Society Pub. Series 2

Yin J., Hou J.L., Boisier, S. et al., 2008, submitted to A\&A.

\end{document}